# dRG-MEC: Decentralized Reinforced Green Offloading for MEC-enabled Cloud Network

Asad Aftab, and Semeen Rehman
Technische Universität Wien, Vienna, Austria
{asad.aftab, semeen.rehman}@tuwien.ac.at

*Abstract*—Multi-access-Mobile Edge Computing (MEC) is a promising solution for computationally demanding rigorous applications, that can meet 6G network service requirements. However, edge servers incur high computation costs during task processing. In this paper, we proposed a technique to minimize the total computation and communication overhead for optimal resource utilization with joint computational offloading that enables a green environment. Our optimization problem is NP-hard; thus, we proposed a decentralized Reinforcement Learning (dRL) approach where we eliminate the problem of dimensionality and over-estimation of the value functions. Compared to baseline schemes our technique achieves a 37.03% reduction in total system costs.

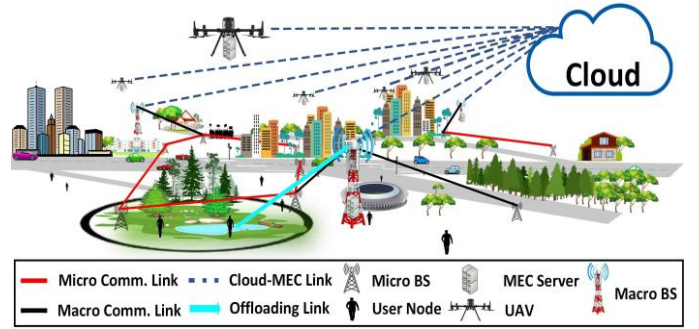

Fig. 1: Our MEC-enabled cloud infrastructure

## I. INTRODUCTION

The rapidly evolving era of mobile communications anticipates 6G wireless networks to deliver unprecedented data generation capabilities. With the increase in the number of mobile devices, requirements such as low latency, energy efficiency, network availability, bandwidth management, and expansive coverage become vital. As wireless networks potentially offer higher bandwidths, the onus also falls on the core network infrastructure to support this advancement.

However, challenges persist. The computation power and battery lifespan of mobile devices struggles to keep pace with the escalating demands of intricate applications that require high data rates with minimum power overhead and latencies. Further, maintaining a high Quality of Service (QoS), which ensures the performance of critical applications regarding latency, throughput, packet loss, availability, etc., within the constraints of network capacity is another essential aspect.

Hence, maintaining high QoS necessitates the use of flexible storage and processing capabilities, like leveraging cloud resources. Yet, due to high power costs, not all applications can connect to cloud data centers, which requires managing cloud resources in a smart manner, where latency-sensitive, real-time applications may be prioritized such as IoT health- care infrastructure, traffic monitoring, etc., where we cannot bear the negligence for data loss [1], [2]. Furthermore, in accordance with the United Nations' Sustainable Development Goal 11, which emphasizes green energy requirements, there is a critical need to evolve computation offloading solutions towards more sustainable green models. This necessitates the importance and requirement of enhancing sustainability within the network infrastructure. In State-of-The-Art (SoTA) [3]–[6] proposes task load management techniques where access to cloud servers is designated for latency-critical applications via a predictive mechanism.

However, load management techniques tend to elevate power consumption and cause over or under-utilization of resources due to prediction rate uncertainties. To overcome these challenges, scholars have proposed solutions centered on the offloading computation of high latency-sensitive applications to multi-access Mobile Edge Computing (MEC) [7], [8]. Recent advancements in MEC, such as the inclusion of High-Performance Servers (HPS), have allowed user nodes to offload data for accelerated computation. Offloading to MEC alleviates network latency, power consumption, and congestion, thus potentially elevating QoS [9], [10].

Yet, for optimized offloading, resource utilization, and better data distribution, an appropriate MEC selection and offloading strategy must be developed. Several researchers have proposed different offloading architectures [11]–[14], primarily based on learning algorithms, to address these issues. For instance, a Semi-Markov Decision Process (SMDP) criterion has been proposed [11], which solves the optimization problem using linear programming for a two-tier data offloading scenario. However, this solution neglects delay constraints and assumes data offloading is only feasible with a nearby cloud network. Researchers in [12] also utilizes SMDP to balance the trade-off between high energy costs and computation availability, employing model-free Reinforcement Learning (RL) methods. However, this architecture only prioritizes high-priority tasks, leading to potential starvation for low-priority tasks.

Furthermore, [13] proposes an RL solution for selecting appropriate collaborative MEC servers and allocating the corresponding portion of the computing task to individual MEC servers and the bandwidth resource, minimizing the average service latency. The maximum tolerable latency is introduced as a constraint. In [14], a dynamic heuristic algorithm is pro- posed that cooperatively allocates computational resources and wireless bandwidth to mobile devices, consequently reducing the overall system's energy.

Since the RL algorithm [15] has been considered a better approach for efficient data offloading [16]. However, with an increased number of agents, the state space increases which leads to a bigger table size resulting in a dimensionality problem in terms of memory usage along with additional delay. In order to mitigate the dimensionality problem, Deep Q Learning (DQL) in [17] has been proposed, resulting in the high scalability of networks. But when it comes to online resource allocation, and scheduling for computation offloading at dynamic MEC positioning [18]–[20], DQL suffers from the problem of overestimation of resource distribution and data offloading.

### A. Problem Motivation and Research Challenge

So far, the state-of-the-art (SoTA) techniques have not adequately addressed the problem of dimensionality and over-estimation of resource allocation and scheduling techniques for offloading tasks at run-time in a dynamic environment of a MEC-enabled cloud network, which leads to unfair resource distribution among different tasks and unbalanced network. Hence, the associated research challenge is, how to develop an optimal resource allocation and scheduling technique such that it minimizes energy, latency, computation, and communication overhead by fairly allocating the resources of MEC nodes among different tasks.

### B. Our Novel Contribution

To address the above-discussed research challenge, we propose a novel dRG-MEC architecture to achieve green resource allocation and computation offloading for MEC. Our dRG-MEC architecture's key attributes are as follows:

- **Joint optimization solution** is proposed for MEC enabled computation offloading, resource allocation, and energy management in a real-time environment. We have formulated this as a Mixed Integer Nonlinear Programming (MINP) problem, under the constraint of limited energy, computation resources, and data offloading capacity. Since the problem has no fixed solution as it is dependent on varying time, energy, latency, and resource distribution constraints, therefore, it is formulated as an NP-hard problem.
- **Decentralized Double Deep Q-Learning** is proposed due to the NP-hard nature of the problem, we have proposed a model-free decentralized Double Deep Q-learning solution that will aid in dimensionality reduction and over-estimation of real-time errors of Q-learning.

The rest of the paper is organized as follows. Section II presents the details of the system modeling, followed by Section III which presents the problem formulation for the joint optimization. Section IV presents the RL solution for green computational offloading and resource allocation problems. The architecture is evaluated and results are presented in Section V. Finally, Section VI provides the conclusion and future work.

## II. SYSTEM MODEL

We consider an integrated MEC-enabled cloud infrastructure as shown in Fig. 1, with M being a set of MEC nodes, deployed in a geographical region with Geo-positioning Loc($m_i$) where $m_i \in M$. Our assumptions are as follows:

- Heterogeneous Geo-graphical area with diverse cellular connectivity is available including, aerial, ground, cooperative, and Device to Device (D2D).
- Edge computing capabilities are available in aerial, ground, and moving MECs. All the MEC nodes have the intelligence to classify the traffic based on the urgency matrix Urg.
- Each MEC node has total computational resources $CR_{max}$.
- For effective use of the channel, Orthogonal Frequency Division Multiplex (OFDM) [21] is utilized to convert all frequency selective channels to flat channels for uplink and downlink that are distantly placed.

### A. Association Model

The MEC $m_i$ connection with node $n_i$ (intends to offload its data) depends on the distance between them, and are mobile in the ecosystem. The position of MEC is denoted by $p_{m_i}(t) = (x_{m_i}(t), y_{m_i}(t), z_{m_i}(t))$, and the node with data to offload position is $p_{n_i}(t) = (x_{n_i}(t), y_{n_i}(t), z_{n_i}(t))$ at a time instance t. Therefore, the distance between them can be expressed as below:

$$D_{m_i,n_i}(t) = (((x_{m_i}(t) - x_{n_i}(t))^2 + ((y_{m_i}(t) - y_{n_i}(t))^2 + ((z_{m_i}(t) - z_{n_i}(t))^2)^{\frac{1}{2}} \quad (1)$$

The path loss (PL) between the MEC $m_i$ and the node $n_i$ depends on propagation environment given by [22]

$$PL_{m_i,n_i} = 20 \log_{10}\left(\frac{4\pi f_c}{c}\right) + 20 \log_{10}(D_{m_i,n_i}(t))L \quad (2)$$

Here, $f_c$: carrier frequency, $c$: speed of light, and $L$: is

$$L = P_{LoS}\eta_{LoS} + (1 - P_{LoS})\eta_{LoS} \quad (3)$$

Here $P_{LoS}$ is the combination of the probability of the line-of-sight link between MEC server $m_i$ and offloading node $n_i$. The aerial MEC $P_{aerialLoS}$ is given by [23]

$$P_{aerialLoS} = \frac{1}{1+\alpha*exp\left(-\beta\left(\frac{180}{\pi}\right)\sin^{-1}\left(\frac{z_{m_i}-z_{n_i}}{D_{m_i,n_i}}\right)-\alpha\right)} \quad (4)$$

Here α, and β are environmental-dependent constants; η presents the additional loss due to free space propagation. While, $P_{groundLoS}$ between ground MEC $m_i$ and user node $n_i$ is given by [24]:

$$P_{groundLoS} = \exp\left(-2r_o\lambda_o \int_0^{D-\frac{\pi}{2}r_o} G(h)dx\right) \quad (5)$$

Here, G(h) is the complimented cumulative distribution function of obstructions, $r_o$ is the mean radius of obstructions, $\lambda_o$ is the density of obstructions in LoS. For an aerial MEC and the user nodes to be in constant motion, it is imperative to assess their QoS constantly for a better association. Hence, pre-association is required which highly depends on the channel gain between the user node and MEC node, which is given by $\arg\max_{m_i} h_{m_i,n_i}(t)$

$$h_{m_i}^{n_i}(t) = \sqrt{\frac{g_{m_i,n_i} \cdot RD_{m_i}^{n_i}(t)}{D_{m_i,n_i} \cdot PL_{m_i,ni}}}, \forall n_i, m_i \quad (6)$$

Here, $g_{m_i,n_i}$ is the channel gain with small scale fading (follows Gaussian distribution) to path loss; $RD_{m_i}^{n_i}(t)$ follows the Rician distribution.

*B. Task Model*

Assuming that each user node $n_i$ has a task to offload to the MEC node $m_i$, that can be characterized after pre-processing the task into five parameters, $\text{task}_{n_i} = (\text{cat}_{n_i}, \text{data}_{n_i}, \text{ck}_{n_i}, \text{th}_{n_i}^{\max}, p_{n_i}(t))$.

Here, $\text{cat}_{n_i}$: categorized task as per $Urg$ e.g., IoT-healthcare data (high priority), weather prediction (low priority), etc., $ck_n$ : computational complexity of the task, $th^{max}$: maximum latency threshold for a specific category of the task, and $\text{data}_{n_i}$: data packet for offloading. We further assume that $\text{task}_{n_i}$ cannot be partitioned due to the low latency requirements. Therefore, it can be either computed locally or offloaded to the nearby available MEC node. The decision depends on the $\text{cat}_{n_i}, \text{data}_{n_i}, \text{ck}_{n_i}, \text{th}_{n_i}^{\max}$ and $p_n(t)$ of the task. Hence, the decision of offloading forms a set which is denoted by Y=$\{\gamma_{1,1}(t), \gamma_{1,2}(t), \ldots, \gamma_{m_i,n_i}(t)\}$, where

$$\gamma_{m_i,n_i}(t) = \begin{cases} 0, & \text{local} \\ 1, & \text{MEC} \end{cases} \quad (7)$$

*1) Time for Execution:* For a $\gamma_{m_i,n_i}(t) = 0$, the execution time for the $task_{n_i}$ is given by:

$$T_{\gamma_{m_i,n_i}(t)=0} = \left(\frac{\text{data}_{n_i} \cdot ck_{n_i} \cdot \mathcal{U}e_{n_i}}{f_{n_i}}\right) \quad (8)$$

Here, $\mathcal{U}e_{n_i}$ is the energy of the user node. While the $\gamma_{m_i,n_i}(t) = 1$, the execution time for the $task_{n_i}$ is as follows:

$$T_{\gamma_{m_i,n_i}(t)\geq 1} = T_{tr,n_i} + T_{m_i} + T_{tr,m_i} + T_{HO} \quad (9)$$

Here, $T_{tr,n_i}$ is the transmission time from $n_i$ to $m_i$; $T_{m_i}$ is the execution time at MEC, and $T_{tr,m_i}$ is the transmission time for sending back results.

During the task transmission, the delays $delay_{tr}$ involved includes uplink, propagation, queuing, and processing delays.

Thus, the total time for the transmission from $n_i$ to $m_i$, execution at $m_i$, and then back to $n_i$ is given as,

$$T_{tr,n_i} = \frac{\text{data}_{n_i} \cdot D_{m_i,n_i} \cdot delay_{tr} \cdot \mathcal{U}e_{tr,n_i}}{R_{m_i}(t) \cdot \lambda_o} \quad (10)$$

$$T_{m_i} = \frac{ck_{n_i} \cdot \text{data}_{n_i} \cdot \mathcal{U}e_{m_i}}{\rho_{m_i,n_i}(t) \cdot \mathcal{F}_{max}} \quad (11)$$

$$T_{tr,m_i} = \frac{cdata_{m_i} \cdot delay_{tr} \cdot \mathcal{U}e_{tr,m_i}}{R_{m_i}(t)} \quad (12)$$

Here, $\mathcal{U}e_{tr,m_i}$ is the Energy to be consumed during transmission, $R_{m_i}(t)$ is the Shannons capacity of channel, and $\rho_{m_i,n_i}(t)$ is the Computing resources proportion at time instance t, $cdata_{m_i}$ is the computed data. If the connection between $n_i$, and $m_i$ is getting weak, the task will be offloaded to the next nearby $m_i$, thus total execution time includes the time for handover, which is given by,

$$T_{HO} = \begin{cases} 0, & h_{m_i}^{n_i}(t) > h_M^{n_i}(t) \\ \|m_{\text{new}} - m_{\text{old}}\| \cdot T_{tr,m_i}, & otherwise \end{cases} \quad (13)$$

*2) Energy Consumption:* For a $\gamma_{m_i,n_i}(t) = 0$, the energy consumption for the $task_n$ is given by:

$$\mathcal{U}e_{\gamma_{m_i,n_i}(t)=0} = f_{n_i} \cdot ck_{n_i} \cdot delay_{process} \cdot \mathcal{P}_{n_i} \quad (14)$$

Moreover, for the $\gamma_{m_i,n_i}(t) = 1$, total energy consumption depends on the transmission (uplink+downlink) and offloading computation that can be given as:

$$\mathcal{U}e_{\gamma_{m_i,n_i}(t)\geq 1} = \mathcal{U}e_{tr,m_i}(t) + \mathcal{U}e_{m_i} + \mathcal{U}e_{tr,n_i}(t) + \mathcal{U}e_{HO}(t) \quad (15)$$

Hence,

$$\mathcal{U}e_{tr,n_i}(t) = D_{m_i,n_i}(t) \cdot delay_{tr} \cdot G(h) \quad (16)$$

$$\mathcal{U}e_{m_i} = \mathcal{F}_{max} \cdot \rho_{m_i,n_i}(t) \cdot \mathcal{P}_{m_i}(t) \quad (17)$$

$$\mathcal{U}e_{tr,m_i}(t) = delay_{tr} \cdot G(h) \cdot D_{m_i,n_i}(t) \quad (18)$$

$$\mathcal{U}e_{HO} = \begin{cases} 0, & h_{m_i}^{n_i}(t) > h_M^{n_i}(t) \\ \|m_{\text{new}} - m_{\text{old}}\| \cdot T_{tr,m_i}, & otherwise \end{cases} \quad (19)$$

*3) Total Cost:* Keeping in consideration the time of execution, and energy consumption, the total cost expressed for $n_i$ to offload data to $m_i$ can be expressed as

$$\mathcal{V}_{\gamma_{m_i,n_i}(t)} = \kappa T_{\gamma_{m_i,n_i}(t)} + (1-\kappa)\mathcal{U}e_{\gamma_{m_i,n_i}(t)} \quad (20)$$

Here $\kappa \in [0,1]$ is the weight-balancing coefficient energy and time of execution. $\kappa$ is used as the minimum energy consumption solution may result in minimum latency, and vice-versa.

III. PROBLEM FORMULATION

The total computational overhead cost of MEC-enabled cloud computation can be expressed as

$$\mathcal{V}_{total}(t) = \gamma_{m_i,n_i}\mathcal{V}_{\gamma_{m_i,n_i}(t)=0} + (1-\gamma_{m_i,n_i})\mathcal{V}_{\gamma_{m_i,n_i}(t)=1} \quad (21)$$

The objective is to lower the total computational overhead cost via a joint optimization of computation offloading, resource

allocation, and energy consumption. We have formulated our optimization problem as:

$$\max_{\Upsilon,\rho,\mathcal{P},\mathcal{U}e} \sum_{t=1}^{T}\sum_{n=1}^{N}\sum_{m=1}^{M} \mathcal{V}_{total}(t)$$

subject to
$C1: \gamma_{m_i,n_i}(t) = \{0,1\}, \quad \forall n_i \in N, m_i \in M,$
$C2: 0 \le \rho_{m_i,n_i}(t) \le 1$
$C3: 0 \le \mathcal{P}_{m_i/n_i}(t) \le \mathcal{P}_{max_{m_i}/max_{n_i}}$
$C4: 0 \le \mathcal{U}e_{\gamma_{m_i,n_i}}(t) \le \mathcal{U}e_{threshold}$
$C5: T_{\gamma_{m_i,n_i}(t)} \le T_{max\_task}, \quad \forall n \in N, m \in M$

The above joint optimization problem can only be solved for a sub-optimal solution where total cost for $n_i$ is maximized during a time instance 't' for $\Upsilon$, $\rho$, $P$, and $\phi$. Power $P(t)$, utilized energy $\mathcal{U}e$, and computational resources propagation rate $\rho_{m_i,n_i}$ are the dynamic variables, while remainders are binary. Thus, the formulated problem is the MINP problem (generally NP-hard). Therefore, a sub-optimal solution for this problem can be extracted through deep learning. While bringing the system to a sub-optimal position and keeping it in the state that it should be, requires the system to be given feedback to rectify itself. Therefore, RL will be utilized to maximize the system staying in an appropriate state.

IV. RG-MEC: REINFORCED GREEN OFFLOADING

In this section, the joint optimization problem is approximated as a Markov Decision Process (MDP). This reduces the overall system computational overhead. The optimization problem is first presented as an MDP based on decentralized RL, and then the problem's optimal solution is achieved through a dDDQL-based algorithm.

*A. decentralized Reinforced Green (dRG)-MEC Modeling*

We have considered an environment where there is multiple $m_i$, and $n_i$ as shown in Fig. 2. The $n_i$ data is either formulated through local computation or offloaded to the $m_i$. Each $n_i$ in the network acts as an agent that interacts with a dynamic environment and experiences different conditions. Then these experiences are distributed through a policy learned through past behaviour.

During a time instance *t*, each agent receives the state $S_{n_i}^t$ of the current environment, and based on learned policy $\pi$ takes action $A_{n_i}^t$. The action of all the nodes forms a joint action $A_N^t$. Such joint action helps the environment to shift to a new state $S_{n_i}^{t+1}$. Afterwards, the agents receive reward $R(t)$ based on their $A_{n_i}^t$. These key elements of dRG-MEC are explained below:

*1) States:* The state space as per our considered environment depends on: the channel condition $h_{m_i}^{n_i}(t)$, transmitted data $data_{n_i}$, node's energy $\mathcal{U}e_{n_i}(t)$, local computational cost $\mathcal{V}_{n_i}^{total}$, data rate available for transmission $R_{n_i}^t$, data categorization complexity $cat_{n_i}$, computational complexity of the task $ck_{n_i}$, and latency threshold $th_{n_i}^{max}$. Moreover, each agent faces a complexity while changing a state and they are highly correlated to training parameters, number of iterations $\Psi$ and the

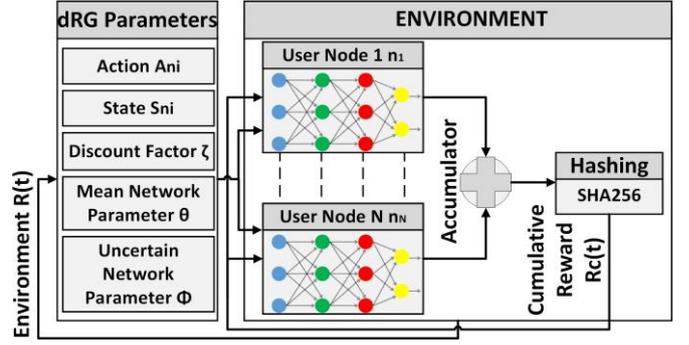

Fig. 2: Our decentralized Reinforced Green (dRG) framework

probability of selection of a random state $\varepsilon$. Thus, $n^{th}$ agent state space will be:

$$S_{n_i}^t = \{h_{m(i)}^{n(i)}(t), data_{n_i}, g_{m_i,n_i}, pos_{m_i}, pos_{n_i}, \mathcal{U}e_{n_i}(t),$$
$$\mathcal{V}_{total}, R_{n_i}(t), cat_{n_i}, th_{n_i}^{max}, ck_{n_i}, \Psi, \quad (23)$$

*2) Actions:* The action space for our proposed architecture consists of six features: sub-band allocation, traffic scheduled $Tr_{n_i}(t)$, data offloading decision $\gamma_{n_i}$, computing resources allocation rate at MEC node $\rho_{m_i,n_i}(t)$, transmission power $\mathcal{P}_{n_i}(t)$, and energy of node $\mathcal{U}e_{tr,n_i}$. For RL, we have discretized the continuous values of power, energy, and computing resources allocation rate. Moreover, $cat_{n_i}, len_{n_i}$ of the $task_{n_i}$ is determined, and then for lowering the complexity, we provide all these parameters to the employed classifier for local vs MEC computation based on the available $\mathcal{U}e$ at the $m_i$ and $n_i$, $th_{n_i}^{max}$, and computation resources available at the $m_i$, and locally at the $n\_\{i\}$.

*3) Rewards:* The reward for the environment in our proposed architecture depends on the total system cost $\mathcal{V}_{total}$. Since in our architecture we are minimizing the offloading computation overhead, energy consumption rate, and latency of the task, thus the reward is negatively correlated to the total system cost. Therefore,

$$\mathcal{R}(t) = \begin{cases} \mathcal{C} - V_{total}, & if\, C1\, to\, C5\, are\, met \\ Penalty, & otherwise \end{cases} \quad (24)$$

Here $\mathcal{C}$ is constant. For a cumulative reward $\mathcal{R}_c(t)$ we define a long-term accumulative reward as:

$$\mathcal{R}_c(t) = \sum_{t=0}^{T} \zeta^t \mathcal{R}(t) \quad (25)$$

Since our proposed model is independent of the $n_i$ and $m_i$ position and the total number of nodes in the overall system, it is highly scalable.

*B. Decentralized Reinforcement Learning*

Based on reward and state transition probability, RL can be either model-based or model-free learning. For online learning, the environment is always non-deterministic, and hence model-free learning is required. Therefore, Q-learning [25] which is one of the model-free learning algorithms is the best candidate for our system model.

```
Algorithm 1 Deep Q-Learning
Initialize: ϵ, ζ, Ψ, θ, N agents
for each training episode η do
    Reset tasks p_{n_i}, environment state, S = S_0
    for each step t do
        for i = 1 to N do
            Observe S^t_{n_i} and sample a
            if sample ≤ ϵ then
                select an action rand (A^t_{n_i} ∈ A^t_N)
            else
                A^t_{n_i} = arg max_{A^t_{n_i} ∈ A} Q*(S^t_{n_i}, A^t_{n_i})
            end
        end
        Action ← Joint Action (A^t_N)
        for i = 1 to N do
            Observe S^{t+1}_{n_i}
            Memory = (S^t_{n_i}, A^t_{n_i}, R(t), S^{t+1}_{n_i})
        end
    end
end
```

```
Algorithm 2 Decentralized Double Deep Q-Learning
Initialize: ϵ, ζ, Ψ, θ, N agents
for each training episode η do
    Run Algorithm 1 Deep Q-Learning
    for i = 1 to N do
        Memory^{(t)}_{n_i} = (S^t_{n_i}, A^t_{n_i}, R(t), S^{t+1}_{n_i})
        min (Loss^t_{n_i})
    end
    θ ← Loss^t_{n_i}
    distribute(H(Memory^{(t)}_{n_i}))
end
```

For resource allocation using Q-learning, an optimal policy can be, $\pi^*$ of each agent. Since DRL utilized the $\epsilon$-greedy approach for balancing exploration ($\epsilon$ for random explore) and exploitation (for exploiting with $1-\epsilon$ chance). In this paper, we use a novel decentralized Double Deep Q-learning (dDDQL)-based approach by taking advantage of a parameterized indexed function for the efficient exploration, where all the agents are decentralized to build a trustful and efficient system.

Every agent in our proposed architecture receives the parameterized state-action value function which is as follows:

$$Q^{\pi^*}_{n_i}(S^t_{n_i}, A^t_{n_i}) = v(\mathcal{R}_c(t)|S^t_{n_i}, A^t_{n_i}) + m(\mathcal{R}_c(t)|S^t_{n_i}, A^t_{n_i}) \quad (26)$$

Here, $v(\mathcal{R}_c(t)|S^t_{n_i}, A^t_{n_i})$, is the mean network for learning and $m(\mathcal{R}_c(t)|S^t_{n_i}, A^t_{n_i})$, an uncertain network for learning with Gaussian random distribution. The $n_i$ agent after taking an action $A^t_{n_i}$ under policy $\pi^*$, has a Q-function as:

$$Q^{\pi^*}_{n_i}(S^t_{n_i}, A^t_{n_i}) = \mathbb{E}\{\mathcal{R}_{t+1} + \zeta \cdot \max_{A^t_{n_i} \in A} Q^{\pi^*}_{n_i}((S^{t+1}_{n_i}, A^{t \to t+1}_{n_i})| S^t_{n_i} = S, A^t_{n_i} = A) \quad (27)$$

This can be simplified by the Bellman optimality equation [26], and will be as below:

$$Q^{\pi^*}_{n_i}(S^t_{n_i}, A^t_{n_i}) = \sum_{S^t_{n_i} \in S} P^a_{S^t_{n_i} \to t+1} (\mathcal{R}_C + \zeta \cdot \max_{A^t_{n_i} \in A} Q^{\pi^*}_{n_i}((S^{t+1}_{n_i}, A^{t \to t+1}_{n_i}))) \quad (28)$$

Here, $\mathcal{R}_c$: Cumulative reward for taking action $A^t_{n_i}$ while going to a state $S^{t+1}_{n_i}$, and $P^a_{S^t_{n_i} \to t+1}$ is the probability of state transition. As Q-learning is deployed for sub-optimal calculation of $Q^{\pi^*}_{n_i}$ by an iteration algorithm that is given as:

$$Q^{\pi^*}_{n_i}(S^t_{n_i}, A^t_{n_i}) \leftarrow Q^{\pi^*}_{n_i}(S^t_{n_i}, A^t_{n_i}) + \Psi[r_{t+1} + \zeta \cdot \max_{A^t_{n_i} \in A} Q^*(S^t_{n_i}, A^t_{n_i}) - Q(S^t_{n_i}, A^t_{n_i})] \quad (29)$$

Here $\Psi$ is the Learning parameter. Let $\theta, \Phi$ be the parameters of the mean and the uncertainty network, and $\theta', \Phi'$ be the parameters for target networks. Thus, the parameters for the mean network are updated through the target mean network and the previous network, as follows:

$$\nabla_{\theta(t)} L_v(\theta(t), \theta(t)', \theta(t-1), d)$$
$$= \frac{1}{|d|} \sum_d \nabla_\theta \left( (v^t_\theta + v^{t-1}_\theta)(S^t_{n_i}, A^t_{n_i}) - Q^{\pi^*}_{n_i}(S^{t-1}_{n_i}, A^{t-1}_{n_i}) \right)^2 \quad (30)$$

Moreover, the updating process for uncertain network parameters is as follows:

$$\nabla_{\Phi(t)} L_m(\Phi(t), \Phi(t)', \Phi(t-1), d)$$
$$= \frac{1}{|d|} \sum_d \nabla_\Phi \left( (m^t_\Phi + v^{t-1}_\Phi)(S^t_{n_i}, A^t_{n_i}) - Q^{\pi^*}_{n_i}(S^{t-1}_{n_i}, A^{t-1}_{n_i}) \right)^2 \quad (31)$$

Here $d$ is delay set $(S^t_{n_i}, A^t_{n_i}, R^t_{n_i}, S^{t \to t+1}_{n_i}, comp)$. The agents learn from previous actions and try to maximize their future rewards, but it may get localized to a small search area, which results in the degradation of the solution. To mitigate such an issue, an epsilon-greedy policy may help as it gives an edge in exploration and exploitation. Exploration is the selection of a random action to see its difference in reward generation from other actions, while action exploitation depends on the prior actions. Thus, the action set for epsilon-greedy will be:

$$\mathcal{A}^t_{n_i} = \begin{cases} rand, & \epsilon \\ arg \max_{A^t_{n_i} \in \mathcal{A}} Q^*(S^t_{n_i}, \mathcal{A}^t_{n_i}), & 1-\epsilon \end{cases} \quad (32)$$

Since the Q-learning in a time-dependent system updates state-value pairs iteratively, that results in extra memory usage. Therefore, Deep Q-Learning (DQL) parameterized by $\theta$ is the suitable candidate for Q-value function approximation. Each agent's weight parameter $\theta$ is updated by a random distribution of a mini-batch of past actions and state experience, in this approximation. Algorithm 1 shows the detailed deep Q learning process for approximation. But, if the aggregation is done at single-agent $n_i$, it will exhaust its memory. Therefore, a new agent is selected for every aggregation and results are distributed after hashing [27].

Also, the Q-learning and Deep Q-learning process tends towards the maximization bias for action evaluation due to the max operator, leading to an over-estimation of some more likely actions to get selected. To evolve from such a problem, a novel decentralized Double Deep Q-learning (dDDQL) is employed which evaluates the next state and the target network will determine the optimal action taken in the next state. Algorithm 2 shows the detailed DDQL process for better action selection. The Target Value Function (TVF) for the DDQL network will become as

$$TVF_{n_i}^t = \mathcal{R}_c(t) + \Psi \cdot \left(S_{n_i}^{t+1}, A_{n_i}^{t+1}; \theta\right) \quad (33)$$

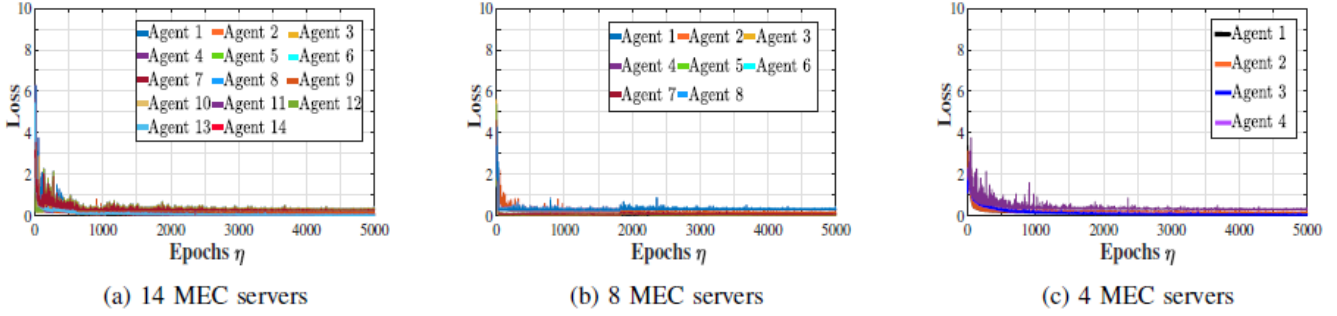

Fig. 3: Loss incurred by dDDQL Reinforcement Learning

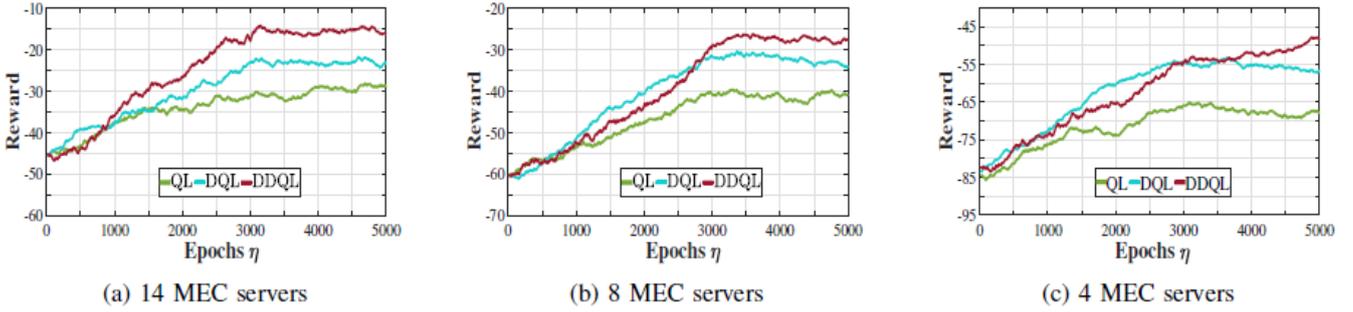

Fig. 4: RL-based Reward Convergence Analysis

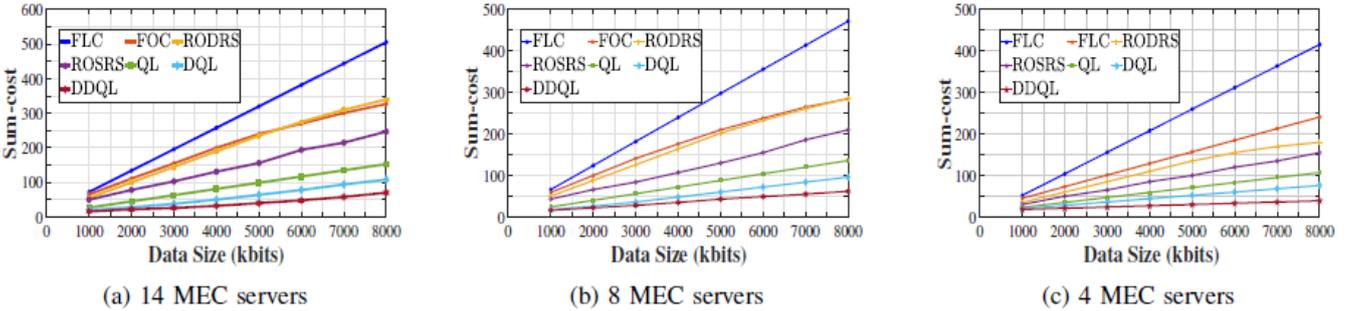

Fig. 5: Total Sum-cost incurred in the system against the data size of tasks to be offloaded

## V. PERFORMANCE ANALYSIS

In this section, we first present the simulation setup that was used to evaluate the performance of the proposed DDQL algorithm. Then the network performance will be examined by comparing our approach with Full Local Computation (FLC), Full Offload Computation (FOC), Random Offloading with Dedicated Resources Strategy (RODRS), Random Offloading and Sharing Resources Strategy (ROSRS), QL algorithm, DQL algorithm, and proposed dDDQL algorithm.

### A. Simulation Setup

We considered a heterogeneous environment as shown in Fig. 1. The edge servers present over moving UAVs, and fixed Base Station (BSs) are initialized with poison distribution for positioning. For training purposes, we had 7 UAVs and 7 fixed BSs MEC nodes in the network. The dDDQL network parameters used for simulating the proposed approach are shown in Table I.

### B. Results

*1) Loss Function*: The loss function values for each MEC agent are analyzed during the training phase as shown in Fig. 3. During the initial training phase, the MEC agents' loss is high because of the huge difference between selected actions, and the resulting Q-values from the targeted Q-values. However, with continuous training, the MEC agents are able to find a sub-optimal solution which results in lowering the overall loss of the dDDQL. Also, due to the reward function, the MEC agents are tending toward the sub-optimal solution in a fast fashion which is evident from Fig. 3.

*2) Convergence*: In Fig. 4, the convergence behaviour of the QL-based, DQQL-based, and proposed multi-agent DDQL-based solutions is shown, which illustrates the cumulative rewards achieved per epoch, for 7 UAV MECs and 7 fixed BSs MECs. It is noticeable that, the total rewards per epoch increase, which evidently shows that the agents are adapting to the environment effectively. Therefore, they are converging as per the sub-optimal policy of joint optimization to maximize the reward. Our proposed DDQL-based algorithm progressively converges around 925, 2710, and 3710 episodes, for 4, 8, and 14 MECs, respectively.

The slight fluctuations are due to the result of the cross-channel interference, channel fading, and cooperative communication conditions superimposed by mobile MECs.

TABLE I: Simulation Parameters

|  | Parameters | Values |
|---|---|---|
| **DDQL Network** | Layers | 6 |
|  | Hidden Layers Neurons | 1000,500,250,120 |
|  | Activation Function | ReLU, Softmax |
|  | Optimizer | Adam |
|  | $\Psi$ | 0.0001 |
|  | $\zeta$ | 0.9 |
|  | $\epsilon$ | $U(1, 0.001)$ |
|  | Epochs ($\eta$) | 5000 |
|  | $t$ | 1000 |
|  | Batch Size | 1500 |
| **Physical Network** | $M$ | 14 |
|  | $N$ | 55 |
|  | $M_{non-fixed}$ CR | $U(0.5, 1.5)$ GHz |
|  | $M_{fixed}$ CR | 6 GHz |
|  | $P_{n_i}$ | $U(-20, 50)$ dBm |
|  | $data_{n_i}$ | [10,80] Mbits |
|  | $ct_{n_i}$ | [1000,5000] Mcycles |
|  | $P_{max_{MEC}}$ | 5 W |
|  | $Tr_{n_i}$ | 0.2 Hz |

*3) Sum Cost Mobile multi-MEC*: The sum cost acquired by the system with respect to the number of mobile multi-MEC in the network is shown in Fig. 6. With the increase of mobile multi-MEC, the total sum cost of the network increases accordingly. The proposed DDQL-based solution outperforms other baseline schemes, indicating a reduction in the computation and communication overhead costs of the network.

Moreover, it can be seen that as the mobile multi-MEC count increases in the system, the FOC method tends towards higher network costs than the FLC method. This is only because the computation resources available at the mobile multi-MEC server exhaust since computation resources are fairly shared. While the DDQL-based solution is in a better place to decide efficiently on computation offloading along with computation resources allocation as per tasks requirement.

*4) Sum Cost incurred by Varying Data Size*: The sum cost acquired by the system with respect to data block size is shown in Fig. 5. With the increase in data size, the total sum cost of the network increases accordingly. The proposed DDQL-based solution outperforms other baseline schemes with a minimal rise in system overall cost. It is only because the computation of large data blocks locally will slow down the system while performing it on the MEC to optimally stabilize the system.

## VI. CONCLUSIONS

In this paper, we proposed a reinforced green computational offloading strategy that is based on the minimization of energy, latency, computational, and communication overhead by fair usage of computational resources of MEC nodes. This minimization problem was formulated as a MINP problem, under resource and tolerable delay constraints. In order to address the NP-hard nature of the problem, we equipped our architecture with a decentralized multi-agent model-free RL technique DDQL that overcomes the problem of dimensionality possessed by QL and the overestimation of DQL. Compared to baseline schemes our technique achieves a 37.03% reduction in total system costs.

In the future, we will be looking forward to the minimization of overall model weights by considering the pruning of RL agents, which will further help in the minimization of the overall system cost. Moreover, the security of the system will be evaluated and HoloBlock [27] will be inculcated into the system.